# Mechanical alignment of suprathermally rotating grains

Alexander Lazarian

Astronomy Department, University of Texas, Austin, TX 78712-1083

## ABSTRACT

It is shown that mechanical alignment of grains can be efficient for grains rotating suprathermally, i.e. with kinetic energy substantially exceeding $k$ (the Boltzmann constant) over any temperature in the system. The paper studies suprathermal rotation caused by $H_2$ formation and the alignment that takes place due to crossover events. Gaseous bombardment in the course of a crossover as well as both gaseous friction and poisoning of active sites are shown to produce alignment. The first type of alignment happens due to the angular momentum deposited by a corpuscular flux with a grain, the second is caused by the change of the mean time back to crossover due to the interaction with a gaseous flux. We show that the two processes act as to decrease the grain cross section in respect to the flux and we find the Rayleigh reduction factor for the joint action of the two processes as well as the range of applicability of each of the processes. Our study indicates that mechanical alignment can be more widely spread than it is generally accepted.

*Subject headings:* dust, extinction — ISM, clouds — ISM, polarization

## 1. Introduction

Although star light polarization by aligned grains was discovered as far back as 1949 (see Hilther 1949, Hall 1949), the cause of grain alignment remains still a bit of a mystery (see Goodman et. al. 1995). The proposed mechanisms can be subdivided into two classes, namely, mechanical and paramagnetic alignment.[1]

At the moment paramagnetic mechanism suggested in its original form by Davis & Greenstein (1951) is believed to be more promising. In fact, its suprathermal modification proposed by Purcell (1975, 1979) and its superparamagnetic or ferromagnetic modification first suggested by Jones & Spitzer (1967) and further developed by Mathis (1986) are widely referred to as the likely candidates for explaining the observed large-scale pattern of polarization. However, the suprathermal mechanism was shown in Spitzer & McGlynn (1979) to produce only a marginal improvement of paramagnetic alignment, while superparamagnetic mechanism was critically discussed by Duley (1978). At the same time, the original Davis & Greenstein proposal was shown to be inadequate in Jones & Spitzer (1967). We believe, that the ambiguous situation with the ISM paramagnetism deserves a special discussion which we started in Lazarian (1995a,b) and intend to continue elsewhere. The present paper is devoted to mechanical alignment.

Mechanical alignment pioneered by Gold (1951, 1952) is believed to be not applicable to the majority of the ISM grains, which according to Purcell (1975, 1979) should rotate with the energies considerably in excess of the thermal one. according to Purcell (1979), the dominant reason for this is a quasi-regular torque due to recoils from nascent $H_2$ molecules being formed at catalytic sites over grain surface. Therefore every suprathermally rotating grain behaves as a tiny gyroscope which tends to preserve its direction of rotation and thus be insensitive to stochastic torques associated with Gold-type processes.[2] However, the very fact that the Gold alignment is not applicable does not necessarily entail that no mechanical alignment is possible at all.

In the present paper we consider two processes that can provide mechanical alignment of grains, which suprathermal rotating arises from $H_2$ formation. The first one is similar to the Gold mechanism in the sense that it is based on depositing of angular momentum by a gaseous flux. The difference is that this deposition takes place during short intervals of crossovers, when the grain angular momentum is minimal. The efficiency of this process depends on the ratio of the kinetic energies of nascent $H_2$ molecules to the kinetic energy of striking atoms and on the accommodation coefficient for atomic hydrogen. In short, if stochastic torques are dominated by $H_2$ formation, the mechanism is suppressed. Contrary to this, the second mechanism does not depend on the stochastic torques arising from the gaseous bombardment, but on the rate at which atoms arrive at grain surface. We show that the frequency of crossovers depend on the orientation of a non-spherical grain in respect to a gaseous flux and this causes alignment. It is likely, that for mildly supersonic drags, the second mechanism prevail; the opposite is true for hypersonic drags. To summarize, our study testifies that suprathermal rotation does not prevent grains from being aligned mechanically and we find that high degree of alignment is attainable.

---

[1]This classification omits a mechanism of ferromagnetic grain alignment suggested in Spitzer & Tukey (1951). However it is shown in Lazarian (1994b) that the latter mechanism can be only important for dark molecular clouds.

[2]Note that to simplify our presentation we do not speak about Harwit (1970) mechanism, which was shown in Purcell & Spitzer (1971) to be inferior to the Gold mechanism for an absolute majority of astrophysically interesting situations.

Below we do not discuss regular torques caused by variations of the accommodation coefficient or photoelectric emission. These processes and the alignment that they can cause are discussed in Lazarian (1994b). Neither we address the important question of whether the mechanical alignment is prevalent within particular regions of the ISM. In our next paper in the series we are going to address this problem by comparing the relative efficiency of different mechanisms for a number of typical ISM regions.

The structure of the paper is as follows. In section 2 we briefly discuss the effect of gaseous bombardment on a suprathermally rotating grain undergoing a crossover, then in section 3 we find analytical expressions for the alignment measure attainable by grain – gas interaction during crossovers. Section 4 deals with the alignment caused by grain – gas interaction during spin-ups. A joint action of the two mechanisms is discussed in Section 5.

## 2. Suprathermal rotation & crossovers

### 2.1. Torques due to $H_2$ formation

Suprathermal rotation of the ISM grains was theoretically discovered by Purcell (1975). There it was shown that the ejection of $H_2$ molecules formed over grain active sites can cause rotation with suprathermal velocities and this should be the major cause of the suprathermal rotation for the ISM grains. Then, in Purcell (1979), it was shown that internal dissipation of energy within grains, mainly due to the Barnett relaxation, suppresses rotation around any axis but the axis of the greatest inertia on the time-scale $\sim \frac{10^7}{\eta}$ s, where $\eta$ is the ratio of grain rotational energy to the equipartition energy $\sim kT$.

This means that each grain rotates around its major axis of inertia that we denote $z$-axis. Therefore the problem becomes one dimensional as the two other components of the torque contribute only to insignificant nutations. The number of $H_2$ molecules ejected per second from an individual site is $\sim \gamma_1 l^2 v_1 n_H \nu^{-1}$, where $\gamma_1$ is the portion of H atoms of number density $n_H$ approach the grain with velocity $v_1$ is reacted to form $H_2$ molecules while $\nu$ is the number of active sites over the grain surface. Then the mean square of a residual torque is

$$\langle [M_z]^2 \rangle \approx \frac{\gamma_1^2}{32} l^6 n_H^2 m_{H_2} v_1^2 E \nu^{-1}, \tag{1}$$

where the coefficient $^1/_4$ accounts for the fact that only components of the angular momentum parallel to the $z$ axis contribute to $M_z$. The mean squared angular velocity $\sqrt{\langle \Omega^2 \rangle}$ of a grain depends on a characteristic time of systematic torques $t_L$ and the frictional damping time

$$t_d \approx 0.6 \frac{m_g}{S_n n m v_1}, \tag{2}$$

where $S_n$ is the grain cross-section for the gaseous flux, $m_g$ is the grain mass, $m$ and $n$ are, respectively, the mass and the number density of gaseous atoms. In fact, it was shown in Purcell (1979, eq.(52)) that

$$\sqrt{\langle \Omega^2 \rangle} = \frac{\langle [M_z]^2 \rangle^{1/2}}{I_z} t_d \sqrt{\frac{t_L}{t_d + t_L}}, \tag{3}$$

where $I_z$ is the $z$ component of the momentum of inertia.

If $t_L \gg t_d$:

$$\sqrt{\langle \Omega^2 \rangle} = \langle [M_z]^2 \rangle^{1/2} \frac{t_d}{I_z}. \tag{4}$$

Assuming that $\nu \approx 100$, $\gamma_1 \approx 0.2$ and $E \approx 0.2$ eV, one obtains $\Omega \approx 10^8$ s$^{-1}$, which considerably exceeds the corresponding thermal velocity of rotation. Further on, for brevity, we will call "suprathermally rotating grains" just "suprathermal grains".

## 2.2. Crossovers

It was shown in Spitzer & McGlynn (1979), that the direction of angular momentum of a suprathermal grain, subjected to a regular torque along its major axis of inertia, does not change in axes $x_0 y_0 z_0$ fixed in space, while its modulus changes as

$$J_{z_0}^2 = L_z^2 t^2 + \text{const}, \tag{5}$$

where $L_z$ is a constant torque along the axis of major inertia. The constant in Eq. (5) is equal to the quadratic sum of $J_x$ and $J_y$:

$$J_\perp^2 = J_x^2 + J_y^2, \tag{6}$$

as this is the value of $J_{z_0}$ at $t = 0$ (where $x$ and $y$ are the body axes of the grain). Therefore the angle between the grain axis and $\mathbf{J}$ direction changes from nearly zero to $\pi$ during a crossover. In other words, during a crossover the direction of $\mathbf{J}$ remains the same, while the rotation in the grain body axes changes from nearly pure rotation about the axis of major inertia to rotation about an orthogonal direction, while the grain gradually flips over. Spitzer & McGlynn called this peculiar period "unclamped time" as during this time $\mathbf{J}$ is not clamped to the major axis of inertia.

Following Spitzer & McGlynn (1979), it is natural to consider dynamical evolution given by Eq. (5) as a zero-order solution of the problem, and the dynamical effects caused by stochastic character of the applied torques as perturbations of the zero order solution. We have seen that the zero-order solution envisages a fixed direction of $\mathbf{J}$ in the inertial $x_0 y_0 z_0$ frame, the deviations from which are due to stochastic torques. In the absence of external gaseous flux all the directions are equivalent and the deviations are isotropic. We will show that in presence of the flux, the deviations become anisotropic and so becomes the density of $\mathbf{J}$ vectors for an ensemble of grains.

In the course of a crossover grain angular momentum approaches zero and therefore grains become very susceptible to stochastic torques. The first kind of these torques is associated with $H_2$ formation and within the adopted model they are isotropic, i.e. mean square increments of the angular momentum $\langle (\triangle J_i)^2 \rangle$ along axis $x_0, y_0, z_0$ are equal. On the contrary, increments $\langle (\triangle J_i)^2 \rangle_{ex}$ associated with flux – grain interaction are anisotropic.

If the flux deposits a fraction $\tilde{\kappa}$ of the squared momentum along the $z_0$ axis, and one may roughly estimate the equilibrium distribution of $\mathbf{J}$ vectors using the formulae

$$\begin{aligned}
\sigma_J^* &\approx \frac{3}{2} \frac{\langle (\triangle J_{z_0})^2 \rangle + \gamma_2^{-1} \langle (\triangle J)^2 \rangle_{ex} \tilde{\kappa}}{3 \langle (\triangle J_{z_0})^2 \rangle + \gamma_2^{-1} \langle (\triangle J)^2 \rangle_{ex}} - \frac{1}{2} \\
&= \frac{3}{2} \left[ \frac{\langle (\triangle J)^2 \rangle_{ex} (\tilde{\kappa} - 1/3)}{3 \gamma_2 \langle (\triangle J_{z_0})^2 \rangle + \langle (\triangle J_z)^2 \rangle_{ex}} \right],
\end{aligned} \tag{7}$$

where $\gamma_2^{-1}$ is the ratio of atoms striking the grain to that included in the $H_2$ molecules formed over grain surface. Note, that for pure hydrogen environment $\gamma_2 = \gamma_1$. The combination $3/2(\tilde{\kappa} - 1/3)$ is the measure of $\mathbf{J}$ alignment in the absence of disturbances caused by torques induced by $H_2$ formation, and $\sigma_J^*$ tends to it if

$$\langle (\triangle J)^2 \rangle_{ex} \gg \gamma_2 \langle (\triangle J_{z_0})^2 \rangle \tag{8}$$

To summarize, if the velocity of bombarding atoms exceeds $(\gamma_2)^{1/2} v_{H2}$, which is of the order of $2.5 \cdot 10^5$ cm s$^{-1}$ for $\gamma_2$ equal 0.2, the collisions dominate in terms of momentum deposited with the grain. In dark clouds, the ratio $\gamma_2$ is much smaller due to the decrease of relative abundance of atomic hydrogen and therefore the importance of grain-atomic collisions increase. In fact, for sufficiently small $\gamma_2$ ratio, gaseous bombardment rather than $H_2$ formation dominates the random torques.[3]

If inequality given by Eq (8) is not satisfied, it is easy to see that the dependence $3/2(\tilde{\kappa} - 1/3)$ is preserved, but the amplitude of alignment becomes $x/(1 + x)$ times smaller, where $x = \langle (\triangle J)^2_{ex} \rangle / (\gamma_1 \langle (\triangle J_{z_0})^2 \rangle)$. If the time between the crossovers is $t_x$, it is possible to show that alignment takes place on the time scale of the order of a few $t_x / \arccos(exp(-F))$, where $F$ is the disorientation parameter introduced in Spitzer & McGlynn (1979) (see also Lazarian (1995a) for an explicit expression).

Accounting for the ISM magnetic field **H** and grain magnetic moment, which is due to the Barnett effect does not alter substantially the picture above. The zero approximation would correspond to the angle between **H** and **J** being preserved; the changes of it due to stochastic torques accounted in the first approximation.

## 3. The measure of alignment

The estimates of the previous section indicate that for rather moderate values of gas-grain velocities a corpuscular flux can provide larger increments of angular momentum than $H_2$ formation. Therefore it is natural to assume that $H_2$ formation causes systematic torques, while stochastic torques are mainly due to a corpuscular flux. In other words, our model enables us to disregard in the first approximation the stochastic contribution caused by $H_2$ formation during the crossovers.

The distribution of angular momentum can be characterized by a function $f(n, \mathbf{J})$, where $n$ is the number of grain-atomic collisions. In general, the direction of **J** should be defined by angles $\theta_1$ and $\varphi_1$ in the "gas reference frame" and by $\theta_2$ and $\varphi_2$ in the "grain reference frame" (see Fig. 1). Henceforth grains will be approximated by spheroids with semiaxes $a$ and $b$.

---

[3]Moreover, the impact of $H_2$ formation can decrease considerably if a substantial part of $H_2$ molecules are formed over sites corresponding to aromatic hydrocarbon, which produces $H_2$ molecules with low kinetic energies (see Duley & Williams 1993, Lazarian 1995a for more details).

As the change of grain angular momentum in the course of an individual collision is small, it is possible to describe the process of **J** alignment by the Fokker-Planck equation (see Reichl 1980, Roberge et al 1993). As a result, following Dolginov & Mytrophanov (1976), we can write

$$\frac{\partial f(\mathbf{x}, n)}{\partial n} = \sum_{i=0}^{2} a_i(\mathbf{x}) \frac{\partial f(\mathbf{x}, n)}{\partial x_i} + \sum_{k,i=0}^{2} b_{ik}(\mathbf{x}) \frac{\partial^2 f(\mathbf{x}, n)}{\partial x_i \partial x_k}, \tag{9}$$

where $x$ is a vector in the phase space with coordinates $J$, $\cos\theta_1$, $cos\theta_2$, $\varphi_1$ $\varphi_2$ and the coefficients $a_i$ and $b_{ik}$ presented in the Appendix A. The solution of Eq. (9) is as follows (see Dolginov & Mytrophanov 1976)

$$f(J, \cos\theta_1, \cos\theta_2, n) = \frac{\text{const}_3}{n^{3/2}} \exp\left(-\frac{J^2(1 + g\cos^2\theta_2 + s\cos^2\theta_1)}{2nb^2p^2(1+s+g)}\right), \tag{10}$$

where $s$ is the external flux anisotropy

$$s = -\frac{1}{2}(\langle p^2 \rangle - 3\langle p_z^2 \rangle)(\langle p^2 \rangle - \langle p_z^2 \rangle)^{-1} \tag{11}$$

and $g$ is the grain non-sphericity

$$g = \frac{1}{2b^2}(a^2 - b^2). \tag{12}$$

Note that $\langle p^2 \rangle$ and $\langle p_z^2 \rangle$ are the averaged squared momentum and its $Z_1$ component (see Fig. 1) transferred to a grain in an individual collision. Both $g$ and $s$ can vary from $-\frac{1}{2}$ to $\infty$. It is easy to see that $g = -0.5$ corresponds to flakes and $g \to \infty$ to needles.

Calculations in Dolginov & Mytrophanov (1976) have shown that the solution given by Eq. (10) is accurate up to $|0.25sg|$ terms for $|s| < 1$ and $|g| < 1$. The accuracy of the solution is of the order $s^{-2}$ for $s \to \infty$ when $|g| < 0$ and $g^{-2}$ for $g \to \infty$ for $s < 0$. If both $g$ and $s$ are large the accuracy is of the order of $g^{-1}$ or $s^{-1}$. To find the distribution function for angular momenta $W(\theta_1, \theta_2)$, one needs to integrate Eq. (10) over the magnitude of angular momentum. This integration provides

$$W(\theta_1, \theta_2) = C(s, g)(1 + s\cos^2\theta_1 + g\cos^2\theta_2)^{-3/2}, \tag{13}$$

where $C(s, g)$ normalizes the distribution so that

$$C(s, g) \int_0^{\pi/2} \mathrm{d}\theta_1 \int_0^{\pi/2} \mathrm{d}\theta_2 \frac{\sin\theta_1 \sin\theta_2}{(1 + s\cos^2\theta_1 + g\cos^2\theta_2)^{3/2}} \equiv 1. \tag{14}$$

A peculiarity of "crossover alignment" as compared to the Gold one (see Lazarian 1994a) is that an elementary process of **J** orientation happens over a short time interval corresponding to the "unclamped time" (see section 2). Grain interaction with the flux results in $\mathbf{J}$ alignment in $x_1 y_1 z_1$ system of reference, which for convenience we will call "gas system of reference". In the gas system of reference orientation of $\mathbf{J}$ is given by $\theta_1$. We remind our reader that during "unclamped time" $\mathbf{J}$ moves in the grain coordinates and this motion corresponds to $\theta_2$ changing from 0 to $\pi$. After a crossover $\mathbf{J}$ stays fixed in the external system of reference but is aligned with the axis of major inertia. For instance, for an oblate spheroidal grain this means that the angle between the grain symmetry axis and the axis of alignment $\mathbf{z}_1$ coincides after the crossover with the angle between $\mathbf{J}$ and $\mathbf{z}_1$. As the time scale associated with crossovers is considerably shorter that that associated with spin-ups, the fraction of grains undergoing crossovers at any given moment is negligible and therefore the measure of $\mathbf{J}$ alignment for oblate grains

$$\sigma_J = \frac{3}{2} \langle \cos^2 \theta_1 \rangle - \frac{1}{2} \qquad (15)$$

coincides for an ensemble of grain with the Rayleigh reduction factor $\sigma$ (see Greenberg 1968). Similarly, for prolate grains $\mathbf{J}$ gets perpendicular to the grain symmetry axis and therefore $\sigma = -0.5\sigma_J$.

The mean value of $\langle \cos^2 \theta_1 \rangle$ can be found using $W(\theta_1, \theta_2)$. Indeed,

$$\langle \cos^2 \theta_1 \rangle = $$
$$C(s, g) \int_0^{\pi/2} \mathrm{d}\theta_1 \int_0^{\pi/2} \mathrm{d}\theta_2 \frac{\cos^2 \theta_1 \sin \theta_1 \sin \theta_2}{(1 + s \cos^2 \theta_1 + g \cos^2 \theta_2)^{3/2}}. \qquad (16)$$

We emphasize, that the alignment happens in two stages, namely, at first $\mathbf{J}$ is aligned in the gas reference frame (during the crossover), i.e. when there is no alignment of $\mathbf{J}$ whatsoever in the grain reference frame, then a perfect alignment of $\mathbf{J}$ is obtained in the grain reference frame (during the spin up).

Although the alignment is attained over a sequence of short time intervals separated by relatively long periods of spin-up this does not make this alignment inefficient. Indeed, during crossovers the grain angular momentum is minimal and therefore it is easy to change its direction. Moreover, in spite of the fact, that the time scale of crossovers might be shorter than the time of grain precession in magnetic field, its lines still represent the axes of alignment for grains. Indeed, at any particular moment different grains of an ensemble undergo different phases of precession.

There exist a considerable difference between the way $\sigma_J$ is obtained in Lazarian (1994a) and the way we find it here. In the former paper it was assumed that the angular momentum is directed along the axis of grain major inertia. Evidently this is not true for the mechanism discussed above. In fact, crossovers are characterized by a disorientation of $\mathbf{J}$ in the grain frame of reference. Therefore according to Eq. (16) we assume a uniform distribution of $\mathbf{J}$ the internal frame of reference as compared to using $\delta$ function distribution for $\theta_2$ assumed in Lazarian (1994a). The calculations relevant to obtaining the analytical expressions for $\langle \cos^2 \theta_1 \rangle$ are performed in Appendix B and the results are shown in Table 1.

Consider at first the upper left-hand corner of Table 1 corresponding to alignment of oblate grains when drift tends to be perpendicular to magnetic field [4] (i.e. $s < 0$ and $g < 0$). For these conditions $\sigma_J$ is shown in Fig. 2. The limiting case $s \to -0.5$, $g \to -0.5$ gives

$$\lim_{\substack{s \to -0.5 \\ g \to -0.5}} \langle \cos^2 \theta_1 \rangle = 2 - \sqrt{2}, \qquad (17)$$

which provides

$$\sigma = \sigma_J = \frac{5 - 3\sqrt{2}}{2} \approx 0.38. \qquad (18)$$

Note, that for the same values of $s$ and $g$ alignment measure $\sigma = 1$ was obtained for thermally rotating grains in Lazarian (1994a). We would like to emphasis that the 3D plot is not symmetrical in respect to interchange of $s$ and $g$. This reflects the asymmetry in the way these two parameters enter the formulae for the alignment measure.

The lower left corner of Table 1 corresponds to oblate grains tending to drift mostly along magnetic field lines (i.e. $s > 0$ and $g < 0$). The relevant measure of alignment is shown in Fig 3.

It is obvious that $s = 0$ (i.e. isotropy of external conditions) should produce no alignment. We will show this for flakes ($g = -0.5$). In this case

$$\begin{aligned}
\langle \cos^2 \theta_1 \rangle &= \lim_{s \to 0} \frac{1}{s} \left( \sqrt{\frac{1}{2}} \frac{\operatorname{arcsinh} \sqrt{2s}}{\operatorname{arctanh} \sqrt{\frac{s}{1+2s}}} \right) \\
&= \lim_{s \to 0} (s\sqrt{s})^{-1} (\sqrt{s} - 1/3 s^{3/2} - \sqrt{s}(1-s) - 1/3 s^{3/2}) = \frac{1}{3}, \qquad (19)
\end{aligned}$$

where series representation of arcsinh $x$ and arctanh $x$ (Granshteyn & Ryzhick 1965, 1.641[2], 1.643[2]) were used. This value of $\langle \cos^2 \theta_1 \rangle$ corresponds to the isotropic distribution of angular momenta and according to Eq. (15) $\sigma_J = \sigma = 0$. The absence of alignment for $s = 0$ is evident from all the figures presented in the paper. As $s$ tends to infinity, $\langle \cos^2 \theta_1 \rangle$ tends to zero and therefore $\sigma_J = \sigma$ tends to $-0.5$ (see also Fig 4). High values of $s$ that are necessary for the alignment are easily attainable when charged grains stream along field lines.

---

[4]When we say "tend to be perpendicular", this means that the direction of the drift is within $[-\arccos 1/\sqrt{3}, \arccos 1/\sqrt{3}]$ if we measure angles from the magnetic field direction. Other angles correspond to the drift which in the adopted terminology " tend to be along" magnetic field. We adopt this terminology just to give a visual picture corresponding to different $s$.

Proceeding with our discussion we consider right upper corner of Table 1 which depicts alignment of prolate grains tending to drift at right angles towards magnetic field lines (i.e. $s < 0$ and $g > 0$). The corresponding $\sigma_J$ is drawn in Fig. 5. It is easy to find that for needles, i.e. $g \to \infty$, $\langle \cos^2 \theta_1 \rangle \approx 0.38$, which means that $\sigma_J \approx 0.08$. Note, that the corresponding Rayleigh reduction factor $\sigma \approx -0.04$. The alignment measure increases for less prolate grains and reaches

$$\sigma_J = \frac{3}{2}\left(2 - \frac{\pi}{2} - \frac{1}{3}\right) \approx 0.14 \qquad (20)$$

for spherical grains. In short, alignment of prolate grains for $s < 0$ is not efficient.

The right lower corner of Table 1 presents alignment of prolate grains tending to drift along magnetic field lines (i.e. $g > 0$, $s > 0$). It is apparent from Fig. 6 that the alignment is efficient for large $s$. The measure for $\mathbf{J}$ vectors tends to $-0.5$ which corresponds to a value of $0.25$ in terms of the measure of axis alignment. Therefore streaming along magnetic field lines should be efficient in aligning prolate grains with their axes along the lines.

To summarize, the alignment discussed above is similar to the Gold one in the sense that it is caused by the momentum deposited by a gaseous flux with the grain. The difference is that the alignment proceeds in short periods of crossovers, when grain is susceptible to the gaseous bombardment. This entails the two stage alignment process discussed above.

All the way above it was assumed that the velocity of gas-grain drift substantially exceed the thermal one. However, it is possible to estimate the influence of thermal motions of atoms. Indeed, in the grain reference frame, these motions are seen as deviations of individual atoms from the mean flux direction. The angle of this deviation can be estimated as

$$\varsigma \approx \arcsin \sqrt{\frac{2\frac{kT}{m}}{u^2 + \frac{kT}{m}}} \qquad (21)$$

and therefore it is possible to obtain the measure of alignment through averaging our results over $\varsigma$. In terms of $s$ parameter used above, the averaging over $\varsigma$ for $\varsigma \ll 1$ corresponds to averaging over

$$\delta s \approx -\sqrt{(1/2 + s)}(2s + 3)\varsigma, \qquad (22)$$

which gives an opportunity to use the graphs above to account for a finite gas – grain drift.

## 4. Alfvénic perturbations & supersonic grain motions

Although above we have discussed alignment for various directions of the grain drift velocity, not all of the corresponding parameters $s$ appear to be equally important when we study grain alignment within the ISM. One of the "chosen" parameters $s \to \infty$ corresponds to the motion of charged grains under the radiation pressure, when this motion is constrained by magnetic field. Estimates in Purcell (1969) show that such a supersonic motion is important in the vicinity of stars, but cannot persist over an appreciable part of the ISM. Therefore a drift under the influence of the perturbed magnetic field will be our major concern below.

First, we study if grains in diffuse clouds decouple from the moving ionized gas at lower frequencies than ions decouple from neutrals and whether this results in the supersonic drift for any conceivable values of the ISM parameters.

It is generally accepted that the Alfvénic supersonic turbulence dominate the dynamics of random motions within the ISM (Aron & Max 1975, Myers 1985, Falgarone & Puget 1986, Elmegreen 1990, Heiles et al. 1992) and waves of high frequency naturally arise due to non-linear steepening of linear polarized and unpolarized waves as they propagate (Elmegreen 1992, McKee et al. 1993). Energy dissipation that accompanies these supersonic motions is believed to be one of the major heating mechanisms for the intercloud (Ferriére et al. 1988) and for interclump (McKee 1989) gas. Here we want to know whether these motions can cause alignment.

If the Alfvénic velocity exceeds that of sound, it is possible to show that the motions of ionized gas are localized mainly in the plane perpendicular to the magnetic field lines (see Alfvén & Fälthmmar 1963) and this corresponds to $s \approx -1/2$. Further on, for the sake of simplicity, we will assume that for Alfvénic perturbations the drift velocity **u** is perpendicular to **H**. This is a good approximation for standing Alfvénic waves. For travelling Alfvénic waves, in general one needs to perform averaging over angles $[\pi/2, \pi/2 - \arctan B_\perp/B_z]$. However we will disregard this within our simplified approach.

Both a mechanical force due to gaseous bombardment and an electromagnetic one act upon a charged grain subjected to the Alfvénic waves. We discuss at first a mechanical force. This force applied to a grain can be expressed as (Kwok 1975)

$$F = S_d n m u \sqrt{u^2 + v_s^2}, \tag{23}$$

where $S_d \approx \pi a^2$ is the grain cross-section[5], $v_s$ is the sound speed, and $u$ is the relative

---

[5]This cross section should be used when the relative velocities are supersonic. If the

gas-grain velocity, which is the difference between the atomic $v_a$ and grain $v_g$ velocities, i.e. $u = v_g - v_a$. For harmonic perturbations $v_a = v_0 \sin \omega_A t$, and the equation of motion can be written as

$$\frac{\mathrm{d}p}{\mathrm{d}\tau} + \mu_1 p \sqrt{\frac{v_s^2}{v_0^2} + p^2} = -\cos \tau, \tag{24}$$

where $\tau = t\omega_A$, and $p = u/v_0$, while $\mu_1 = \frac{mnS_d v_0}{m_g \omega_A} \approx \frac{3mn}{4\varrho_g} \frac{v_0}{a \omega_A}$ [6], where $\varrho_g$ is the grain density, $a$ is the grain radius, and $\omega_A$ is the Alfvénic frequency and we remind the reader, that $m_g$ and $m$ are, respectively, the masses of a grain and an atom. For subsonic motions, the solution is

$$u = \frac{v_0}{\sqrt{1 + \mu_1^2}} \sin(\omega_A t + \phi_{sh}), \tag{25}$$

where $\tan \phi_{sh} = \mu_1$. This solution reflects the most essential features of the generalized problem, particularly, a decrease of the amplitude of $u$ with $\mu_1$. The lower limit for this coefficient can be expressed as (Lazarian 1994a)

$$\mu_1^{min} \approx i_*^{-1} \frac{3mv_s}{8\rho_g a \langle \sigma_T v_T \rangle}, \tag{26}$$

where $i_*$ is the ionization ratio, $m_a$ is atomic mass, $\langle \sigma_T v_T \rangle \approx 1.5 \cdot 10^{-9}$ cm$^3$ s$^{-1}$ is the collision rate coefficient (Nakano 1984). To obtain the above estimate we used the maximum frequency of Alfvénic waves that still move ionized and neutral components together $\omega_{max} \approx 2i_* n \langle \sigma_T v_T \rangle$ (McKee et al 1993).

Substituting $i_* = 10^{-4}$, $n = 15$ cm$^{-3}$, $a = 10^{-5}$ cm and $\rho_g = 3$ g cm$^{37}$, one obtains $\mu_1^{min} \approx 2 \cdot 10^{-2} \ll 1$, which indicates that supersonic Alfvénic waves should produce supersonic grain drift. The corresponding cut-off frequency $\omega_{max} \approx 4.5 \times 10^{-12}$ s$^{-1}$. Photoionization can increase the ionization ratio (McKee 1989) and data in Myers & Khersonsky (1995) indicate, that the ionization ratio is high up to the densities corresponding to dark clouds.

For our estimates above we have assumed that a grain is a sphere. Our computations above show that under gaseous bombardment grains tend to align in such a way as to

---

velocities are subsonic, one should use the total surface of grains, but the flux becomes $0.25\varrho v_s$. For a sphere, the force is well approximated by $\pi a^2 \varrho u \sqrt{u^2 + v_s^2}$ for both supersonic and subsonic motions.

[6]Note, that $\mu_1$ is proportional to the ratio of the Alfvénic wave period to the the time $t_d$.

[7]Following Spitzer (1978) we assume that $\rho_g = 3$ g cm$^3$ for $a = 10^{-5}$ cm and decreases to $\approx 1$ g cm$^3$ for grains of radius $5 \times 10^{-5}$ cm.

minimize their cross-section. Therefore a further decrease of the upper limit of $\mu_1^{min}$ is expected for non-spherical grains.

In short, for grains with negligible charge, supersonic drift must be widely spread in the ISM for grains of radii above $10^{-6}$ cm. Charging of grains modifies this conclusion. In fact, the charge issue is a subtle topic as the two main causes, i.e. collisions and photoelectric emission, result in charges of opposite signs and the grain charge in diffuse clouds depends on the interplay of these two processes. To start with, consider the charge caused by a disparity in the electron and ion collisional rates. According to Spitzer (1978) the grain potential energy $eU$ becomes of the order of $-2.5kT$, where $T$ is the ion kinetic temperature. A grain with such a charge moves about magnetic field lines with a cyclotronic frequency

$$\omega_c = \frac{UaB}{m_g c} \approx \frac{2.5kTaB}{em_g c}.$$

(27)

For $\rho_g = 3$ g cm$^{-3}$, $a = 10^{-5}$ cm, $T = 80$ K and $B = 3 \times 10^{-6}$ G Eq. (27) provides $\omega_c \approx 4.6 \times 10^{-12}$ s$^{-1}$, which is of the same order that $\omega_{max}$. The ratio $\mu_2 = \omega_c/\omega_A$ is another important parameter of the theory for $\mu_2 < 1$ the drift velocity scales approximately as $v_0\mu_2$. This limits the sizes of grains that move supersonically. In fact, only grains with radii above $10^{-5}$ cm are likely to drift supersonically as a result Alfvénic oscillations, if the values $\omega_c$ and $\omega_{max}$ are given as above. Incidentally this corresponds to the size distribution of grains that cause polarization (see Kim & Martin 1994, 1995). If the charge is reduced by photoelectric emission or $\omega_{max}$ is greater, the supersonic drift can persist in the diffuse clouds for smaller grains.

In short, in order to provide supersonic drift of grains of the size $\sim 10^{-5}$ cm[8], high frequency Alfvénic waves should have large amplitudes. This does not contradict to our present day knowledge of Alfvénic waves, which "evolve by steepening until ion-neutral collisions damp them" (McKee et al 1993) and cannot be dismissed by considering the energy balance within galactic interstellar matter either. However more elaborate study is needed before any definite far-reaching conclusion can be made.

Grain charge provides a possibility for alignment through supersonic ambipolar diffusion. This process cannot be ubiquitous diffuse clouds as an assumption of supersonic ambipolar diffusion over large scales entails energy dissipation well in excess of what supernova explosions can inject into the ISM turbulence. However, this may be an option for particular regions subjected, for instance, to MHD shocks. Note, that the importance of ambipolar diffusion in the context of mechanical alignment of grains was first mentioned in Whittet (1992) with the reference to a forthcoming paper by Roberge & Hanany.

---

[8]If small grains drift with supersonic velocities, then a fortiori larger grains should drift supersonically.

According to Draine, Roberge & Dalgarno (1983), grain velocity in respect to neutral gas is

$$u = (v_n - v_i)\frac{\omega_c t_m}{\sqrt{1 + (\omega_c t_m)^2}}, \tag{28}$$

where $v_n$ and $v_i$ are the velocities of neutral and ionized components respectively, while $\omega_c$ is the grain gyrofrequency and $t_m$ is the time that takes a grain to collide with gaseous atoms of the net mass equal to that of the grain. According to fig. 1 in Pilipp $et$ $al.$ (1990), grain-neutral velocities may be highly supersonic for sufficiently strong shocks.[9]

Hereafter while discussing alignment under Alfvénic perturbations we will bear in mind both supersonic grain drifts in highly ionized media due to Alfvénic waves as well as grain drift due to magnetohydrodynamic shocks in weakly ionized media.

The measure of alignment for oblate and prolate grains under Alfvénic perturbations is shown in Fig 7 and Fig. 8 respectively. It is obvious that for prolate grains the alignment is negligible. At the same time a comparison between fig. 5a in Lazarian (1994a) and Fig. 7 indicates that although the alignment for flakes is less efficient for suprathermally rotating grains as compared with thermally rotating grains, for grains of moderate oblateness, both processes deliver a comparable degree of alignment, which is of the order of 20%.

## 5. Alignment in the sequence of crossovers

Suprathermal grains are not sensitive to the angular momentum deposited by chaotic gaseous bombardment during spin-ups. However, this does not mean that gaseous flux does not cause alignment of suprathermal grains apart from relatively short crossover intervals.

We will show below that the gaseous flux influences the distribution of the angular momentum for an ensemble of non-spherical suprathermal grains. The condition of non-sphericity is essential for the mechanism under study, as we will show that it is the difference in cross section of the flux - grain interaction that produces the anisotropy. For spherical grains, which are the favorite object for theoretical studies, the effect vanishes. In other words, as the angular momentum deposited with the grain by the gaseous flux becomes not important as compared with the overall angular momentum of the grain, it is the rate at which atoms arrive to the grain surface that matters. This rate influences both the diffusion of the $\mathbf{J}$ and the frequency of crossovers; we will show further on that either of these effects is capable to produce an anisotropic distribution of grain axes.

---

[9]Note, that if the charge density of grains is a significant fraction of that of the ions and electrons the grain drag is independent of the parameter $\omega_c t_m$ (Nakano & Umebayashi 1980).

Due to stochastic torques the direction of $\mathbf{J}$ is subjected to diffusion, the characteristic time of which can be estimated as the time during which the mean square deviation $\sqrt{\triangle J_\Sigma^2}$ becomes of the order of $\sqrt{\langle J^2 \rangle}$. This diffusion is a random walk process and therefore:

$$\triangle J_\Sigma^2 \approx \triangle J_{el}^2 \frac{t}{t_{el}},$$ (29)

where $\triangle J_{el}$ is the angular momentum deposited in an individual elementary torque event and $t_{el}$ is the mean time between the torque events. As, according to Eq (3), the mean squared angular momentum is

$$\sqrt{\langle J^2 \rangle} \approx \nu^{-1/2} \triangle J_{el} \frac{t_d}{t_{el}} \sqrt{\frac{t_L}{(t_d + t_L)}}$$ (30)

the characteristic time of diffusion is

$$t_{diff} \approx \nu^{-1} t_d \frac{t_d}{t_{el}} \frac{t_L}{t_d + t_L}.$$ (31)

In general, both $\triangle J_{el}$ and $t_{el}$ in Eqs. (29) and (30) are different. For instance, if stochastic torques caused by atomic bombardment dominate, $\triangle J_{el}$ and $t_{el}^{-1}$ in Eq (29) are, respectively, angular momentum and frequency associated with atomic impacts; those quantities can differ several times from ones entering Eq (30). However, for our order of magnitude estimates this should not be very important.

Consider now crossovers. From the point of view of $\mathbf{J}$ dynamics in angular coordinates they are equivalent to leaps; the corresponding time of disorientation as a result of such leaps is

$$t_{leap} \approx \frac{t_x}{\arccos(\exp(-F))},$$ (32)

where we remind the reader, that $F$ is the disorientation parameter. For sufficiently small grains $\arccos(e^{-F}) \approx \pi/2$, which corresponds to a complete disorientation during a crossover. The time $t_x$ was estimated in Purcell (1979) as

$$t_x = 1.3(t_L + t_d).$$ (33)

It is easy to see that $t_{diff}$ equals to $t_{leap}$ if

$$t_L = -t_d + \frac{t_d^2}{2.6\nu t_{el}} \pm \frac{t_d^2}{2.6\nu t_{el}} \sqrt{\frac{4\nu t_{el}}{1.3 t_d} + 1}.$$ (34)

For an ordinary grain $t_d/t_{el}$ is of the order $10^9$, which is of the order of the atomic to grain mass ratio, while $\nu$ does not exceed $10^6$ even for the surfaces completely covered by active sites. Therefore one may expand the square root in Eq (34) and to obtain, that for the diffusion to dominate $\mathbf{J}$ dynamics either

$$t_L \gtrsim t_d \frac{t_d}{\nu t_{el}} \tag{35}$$

or

$$t_L \lesssim t_{el}\nu. \tag{36}$$

The condition given by Eq (35) is not likely to be satisfied for the typical ISM conditions (see Spitzer & McGlynn 1979). Moreover this situation is not interesting from the point of view of mechanical alignment, as the expected paramagnetic alignment is nearly complete for such long $t_L$.

The condition given by Eq (36) is, in fact, a criterion for the existence of quasi-regular torques[10]. Any H atom arriving at the grain surface has nearly equal chances to form an $H_2$ molecule over any of $\nu$ active sites. As these sites are expected to be distributed randomly over grain surface, the grain experiences random torque over the interval less than $t_{el}\nu$. Such short $t_L$ are expected for grains less than a critical size (Lazarian 1995a,b). For grains smaller than this size, every oxygen atom has good chances to poison an active site. As the poisoning time of $\nu$ active sites is inversely proportional to the oxygen flux, the above criterion (see Eq (36)) for $l < l_{cr}$ is equivalent $n_H\gamma_1/n_o < 1$, where $n_o$ and $n_H$ are the concentrations of atomic oxygen and hydrogen respectively. Therefore, it is obvious, that for reasonable values of the accommodation coefficient $\gamma_1$, e.g. $\gamma_1 = 0.2$, diffusion can dominate leaps only for the core regions of molecular clouds. Above it was implicitly assumed that in spite of rapid poisoning there exist an efficient mechanism of creating new sites. In fact, $H_2$ formation is likely to be suppressed for $l < l_{cr}$. In short, for the majority of cases disorientation during crossovers dominates the dynamics of $\mathbf{J}$.

To find the angular distribution of $\mathbf{J}$, we need to consider a long ($t \to \infty$) sequence of individual "leaps". We assume that in the course of a sequence of "leaps" $\mathbf{J}$ has an equal probability of obtaining any direction within $[0, 2\pi]$. In this case, if $\mathbf{J}$ spends time $t_j$ in a particular volume of the phase space, which is a sum of the $\triangle t_i$ intervals that it spends any time on entering the volume, $\lim_{t\to\infty} \frac{t_j}{t}$ gives the time averaged probability of $\mathbf{J}$ entering the volume. According to the ergodic hypothesis this average coincides with an ensemble one. This way of reasoning is easy to generalize to a continues distribution.

---

[10]The diffusion can be prevalent over leaps in spite of the existence of quasi-regular torques, if the random torques are dominated by gaseous bombardment.

In our two dimensional space of angular coordinates, the time intervals $\triangle t_j$ on average are proportional to $t_x$. The damping time $t_d$ that enters the expression of $t_x$ is inversely proportional to the rate at which atoms arrive to the grain surface (see Eq.(2)); it also natural to assume that $t_L$ varies in the same proportion. Indeed, the time $t_L$ is proportional to the time of colliding with $N_1$ heavy atoms which will poison at least half of active sites; the rate of heavy atoms arriving to the surface is proportional to $S_n$. Therefore $t_{jL}$ and $t_{jx}$, which are, respectively, the life-time of Purcell's rockets and the mean time between crossovers for the position $j$, are inversely proportional to the corresponding cross section $S_{jn}$; the coefficient of proportionality is the problem of normalization. Thus a statistical description of the alignment of a suprathermally rotating grain subjected to a supersonic flow becomes a matter of determining how grain-gas cross section changes in the course of grain precession in the ambient magnetic field. This problem can be solved for grains of arbitrary shapes.

For diffusion of $\mathbf{J}$ when $t_{el} < t_{el}\nu$, it is possible to see that $J$ scales as $\sqrt{t_d} \sim S_n^{-1/2}$ (see Eq (30)), while the deviation $\triangle J$ scales as a square root of the number of torque events, i.e. $\sim S_n^{1/2}$. Therefore the time $\triangle t_i$ and $t_j$ scales again as $S_n^{-1}$. This universality of scaling for "leaps" and "diffusion" regimes is a consequence of the fact, that both $\triangle J$ and $J$ are controlled by the intensity of the flux. If we assume that $\triangle J$ is controlled by some other process, e.g. by cosmic ray bombardment(see Salpeter & Wickramasinghe 1969, Sorrell 1995), while $J$ is limited by friction caused by a supersonic flux, $t_j$ would scale as $S_n^{-1/2}$. We briefly discuss this alignment in Appendix C.

To simplify our treatment, while elucidating the nature of the effects, consider grains approximated by infinitely thin discs; a quantitative treatment of alignment for other shapes will be given elsewhere. An advantage of using a disc, rather than other shape is that a straightforward expression is available for $S_n$:

$$S_n = \pi r^2 \cos \alpha, \tag{37}$$

where $\alpha$ is the angle between the direction of the flux defined as the direction of $\mathbf{u}$ and that of $\mathbf{J}$ ($\mathbf{J}$ is perpendicular to the disc plane as a result of the Barnett relaxation). If $\mathbf{J}$ and $\mathbf{u}$ make, respectively, angles $\varphi$ and $\phi$ with $\mathbf{H}$ and the angle between the planes $\mathbf{uH}$ and $\mathbf{JH}$ is $\psi$ it is obvious from the spherical trigonometry (see Fig 9) that

$$\cos \alpha = \cos \phi \cos \varphi + \sin \phi \sin \varphi \cos \psi. \tag{38}$$

The distribution function $f$ should be averaged over $\psi$ as this is the azimuthal angle of $\mathbf{J}$ that changes in the course of precession about $\mathbf{H}$ direction. Therefore the distribution function

$$f(\varphi, \phi) = C^{-1} \int_0^{2\pi} \frac{d\psi}{\cos \alpha}, \tag{39}$$

where

$$C = \int_0^{2\pi} d\psi \int_0^{\pi} \frac{\sin \varphi d\varphi}{\cos \alpha}, \tag{40}$$

provides an opportunity to obtain the Rayleigh reduction factor for any $\phi$ through numerical integration:

$$\sigma = \frac{3}{2} \int_0^{\pi} \cos^2 \varphi f(\varphi, \phi) \sin \varphi d\varphi - \frac{1}{2}. \tag{41}$$

However, for the most important case, corresponding to Alfvénic perturbations, all the calculations can be done analytically and $\sigma = 0.25$. Such an alignment, although less efficient than the one we discussed in details in section 3 cannot be ignored. For streaming along magnetic field lines $\sigma = -0.5$. A joint action of the two mechanisms is discussed below.

Note, that the alignment caused by Alfvénic perturbations is marginal for grains rotating due to cosmic ray bombardment (i.e. $\sigma = 0.04$). This does not mean, nevertheless, that the mechanism is not efficient at all. For instance, if the flow is directed along magnetic lines, it is possible to show that the corresponding measure of alignment $\sigma$ is equal to 0.2. However, it is beyond the scope of the present paper to discuss this effect.

It worth noting, that the possibility of alignment of grains subjected both to supersonic drift and to cosmic ray bombardment was mentioned back in Salpeter & Wickramasinghe (1969). However, the authors believed that the alignment would be orthogonal to that attainable through the Gold mechanism. On the contrary, we have proved that for non-spherical grains the two mechanisms act in the same direction. This was not found in Purcell & Spitzer (1971) as there spherical grains were studied, while the effect we speak here is caused by grain non-sphericity. The residual marginal alignment obtained in the latter paper arizes from the assumed peculiar interaction of atoms with a grain. Namely, in the model adopted in Purcell & Spitzer (1971) colliding atoms were not allowed to diffuse over grain surface. This poses an interesting question to what extend our results depend on the assumption of atoms being adsorbed by the surface on the collision. For spherical grains Purcell & Spitzer found that the gaseous friction for the rotational axis perpendicular to the flow exceeds 1.5 times that for the rotational axis parallel to the flow. This friction influences $t_d$ and therefore for $t_L \ll t_d$ the effect of atoms evaporating from the same spots over that they hit the grain surface may act in the opposite direction as compared to the effects of crossover difference studied above. However, it is easy to see that for the Alfvénic perturbations the Purcell & Spitzer effect is suppressed due to grain precession. Moreover, for sufficiently non-spherical grains, which are most interesting as the media responsible for dichroic absorption, the alignment is dominated by the difference in cross section, that is studied above. As soon as $t_L$ becomes comparable to $t_d$ the Purcell & Spitzer effect becomes negligible even for spherical grains.

However, the fact, that in our simplified model we may ignore "localized" adsorption – desorption events does not mean they do not deserve a further study. Note, that for most favorable conditions, e.g. when the flow is parallel to magnetic field lines, such "localized" events may provide alignment measure of **J** for spherical grains of the order of $5/17 \approx 1/3$. This estimate is greater that that in Purcell & Spitzer (1971) as we considered quasi-regular torques as opposed to random torques in the latter study. However, a more detailed discussion of the consequences of "localized" adsorption – desorption events is beyond the scope of the present paper.

Above grains were approximated by discs. Another extreme corresponds to needles. It is easy to show that the measure of alignment for them tends to zero for the fluxes perpendicular to magnetic field lines and approaches maximum of the order of 0.1 for fluxes along magnetic field lines. This apparent inefficiency of alignment for prolate grains as compared with oblate ones stems from a marginal difference in the time averaged grain - gas cross sections for different orientations of rotating prolate grain. Therefore in major cases it is possible to disregard the contribution from mechanically aligned suprathermal prolate grains. Note that there is an observational evidence that the aligned grains are typically oblate (see Hindelbrand 1988).

## 6. Discussion

We have seen that suprathermal grains can be aligned by a supersonic gaseous flux. It is also shown, that both mechanisms discussed tend to minimize the cross section of the flux interaction with the suprathermally rotating non-spherical grain.[11]

As the alignment depends on the relative gas-grain velocity (see section 3), one may predict that larger grains in diffuse clouds should be more efficiently aligned by Alfvénic waves, as compared to smaller ones. This is a tendency that corresponds to observations (Mathis 1979). However, drift velocities caused by ambipolar diffusion do not decrease for small grains. Therefore if ambipolar diffusion is responsible for alignment of grains in dense clouds, small grains should be also well aligned. If small grains are non-spherical, this trend may be detectable. For instance, it was suggested by B. Draine (private communication) that PAH might be aligned by the ambipolar diffusion and this may produce polarization in the associated emission features.

---

[11]Note, that the Gold mechanism acts in the same way.

Everywhere above crossover and spin-up alignments were discussed separately. In fact, they act together and we may estimate the resulting Rayleigh reduction factor. In general, if the component of $\mathbf{J}$ along magnetic field is $J_z$, the alignment due to a mechanism acting alone results in the ratio $x_i = \frac{\langle J_{x(i)}^2 \rangle}{\langle J_{z(i)}^2 \rangle} = \frac{\langle J_{y(i)}^2 \rangle}{\langle J_{z(i)}^2 \rangle}$ and the measure of alignment for this particular process can be estimated as follows

$$\sigma_J^{(i)} \approx \frac{3}{2} \left( \frac{1}{2x_i + 1} - \frac{1}{3} \right).$$  (42)

If $m$ independent processes act simultaneously, the corresponding ratio

$$x_\Sigma = \frac{\langle J_{x(\Sigma)}^2 \rangle}{\langle J_{z(\Sigma)}^2 \rangle} = \frac{\langle J_{y(\Sigma)}^2 \rangle}{\langle J_{z(\Sigma)}^2 \rangle} \approx \Pi_{i=1}^m x_i$$  (43)

determines $\sigma_J^{(\Sigma)}$. This gives a way of expressing $\sigma_J^{(\Sigma)}$ through $\sigma_J^{(i)}$

For example, if the crossover alignment corresponds to $\sigma_J^{(1)}$ and the alignment in the sequence of crossovers corresponds to $\sigma_J^{(2)}$ their joint action corresponds to

$$\sigma_J^{(\Sigma)} \approx \frac{\sigma_J^{(1)} + \sigma_J^{(1)} \sigma_J^{(2)} + \sigma_J^{(2)}}{1 + 2\sigma_J^{(1)} \sigma_J^{(2)}}.$$  (44)

Thus, the measure of alignment for flakes subjected to Alfvénic perturbations can be estimated $\approx 0.61$ if the two mechanisms act together.

If, one of the processes dominates and, for instance, $\sigma_J^{(1)} \gg \sigma_J^{(2)}$ it is easy to see that

$$\sigma_J^{(\Sigma)} \approx \sigma_J^{(1)} + \sigma_J^{(2)}(1 - \sigma_J^{(1)}).$$  (45)

If recoils from the $H_2$ molecules formed over grain surfaces dominate the random torques, the alignment in the course of crossovers is suppressed and grains can be mechanically aligned only by the second mechanism. This is likely to be true for relative velocities $u$ just above the sonic ones or if the accommodation coefficient is greater than we assumed. Alternatively, if $t_L > t_d$ and, due to some reason, $t_L$ is independent of the accretion rate, it is the first mechanism that operates.

In fact, our study also indicates the existence of another type of alignment mechanism. Namely, if $t_L$ is controlled by photodesorbtion and $t_L \gtrsim t_d$, the distribution of $\mathbf{J}$ will become anisotropic even in the absence of any mechanical flows; grains will tend to align with their axes of major inertia perpendicular to the direction towards the UV source. For $t_L \gg t_d$, $t_x$ which is $\sim t_L$ becomes again proportional to $S_n^{-1}$ and $\sigma$ can be computed in accordance with Eq.(41). The mechanism will tend to align long axis of grains parallel as well as perpendicular to magnetic field lines depending on the angle between the direction of magnetic field and that towards the UV source. For typical ISM parameters, one needs to account for paramagnetic relaxation if $t_L \gg t_d$ and therefore the resulting alignment can be estimated according to Eq.(44), if one denotes $\sigma_J^{(2)}$ the measure of paramagnetic alignment. However, a detailed discussion of this radiation driven mechanism is far beyond the scope of our present paper.

Everywhere above we disregarded paramagnetic relaxation of the suprathermally rotating grains. This is justifiable if we adopt the "standard values" of the ISM magnetic field and gaseous density and assume $t_L < t_d$ (see Spitzer 1978). If it is not the case, one may use Eq. (44) to estimate the alignment for the joint action of mechanical and paramagnetic processes.

## 7. Conclusions

We have shown that suprathermal rotation caused by $H_2$ formation does not prevent grains from being aligned mechanically. The alignment arizes from both gas depositing momentum with grain during crossovers and due to the change of the mean time back to crossover. These two processes act in the same direction and tend to minimize the cross section of grain interaction with a gaseous flux; the second insures that grains exhibit alignment even if the random torques are dominated by $H_2$ formation. Alternatively, if the gaseous bombardment dominates random torques, the first process dominates. Our results show that the alignment is efficient for oblate grains subjected to Alfvénic perturbations or radiative fluxes. They also indicate that the alignment caused by Alfvénic waves can be widely spread.

This work was initiated by Bruce Draine's comments, but would not be possible if not for encouragement by Ethan Vishniac. I am grateful to Russell Kulsrud for valuable comments on grain charge, Alyssa Goodman and Phil Myers for illuminating discussions on observational data, and to David Williams for explaining me the most subtle issues of grain chemistry. The manuscript was revised after my visit to Princeton University Observatory where I got much from stimulating discussions, especially, with Bruce Draine and Lyman Spitzer. It is a pleasant debt to thank Jeremiah Ostriker for arranging financial support for this visit. This paper also owes much to my communications with John Mathis, Martin

Rees and Paul Shapiro. The research is supported by NASA grant NAG5 2773.

## A.    Coefficients of the Fokker-Planck equation

The coefficients $a_i$ and $b_{ik}$ of the Fokker-Planck equation are determined by the change of the angular momentum of the grain due to its collision with an atom:

$$a_0 = \left\langle \frac{1}{2} \triangle x_2 \frac{\partial \triangle x_1}{\partial x_2} + \frac{1}{2} \triangle \varphi_1 \frac{\partial \triangle x_1}{\partial \varphi_1} - \triangle x_1 \right\rangle \tag{A1}$$

$$a_m = \left\langle \frac{1}{2} \triangle x_1 \frac{\partial \triangle x_m}{\partial x_1} + \frac{1}{2} \triangle x_m \frac{\partial \triangle x_m}{\partial x_m} \right.$$
$$+ \left. \frac{1}{2} \triangle \varphi_m \frac{\partial \triangle x_m}{\partial \triangle \varphi_m} - \triangle x_m \right\rangle \tag{A2}$$

$$b_{ik} = \langle \triangle x_i \triangle x_k \rangle, \tag{A3}$$

where $m = 1, 2$,   $k, i = 0, 1, 2$, and the angular brackets denote averaging over the impacts of atoms over the grain surface. The quantities $\triangle x_i$ and $\triangle \varphi_i$ and the corresponding $a_i$ and $b_{ik}$ were calculated in Dolginov & Mytrophanov [1976] using the following equations:

$$\triangle x_j = \triangle(\mathbf{e}_j \cdot \mathbf{J}) - (\mathbf{e}_j \cdot \mathbf{J}) \triangle J$$
$$\triangle x_0 = \mathbf{j} \cdot \triangle \mathbf{J}$$
$$\triangle \varphi_j = ((\mathbf{e}_j \times \mathbf{j}) \cdot \triangle \mathbf{J}(J(1 - x_j^2))^{-1},$$

where $j = 1, 2$, $\mathbf{e}_1 = \frac{\mathbf{H}}{|\mathbf{H}|}$ is a unit vector along the magnetic field, $\mathbf{e}_2 = \frac{\mathbf{a}}{|\mathbf{a}|}$ is a unit vector along $Z_1$-axis of the grain, $\mathbf{j} = \frac{|\mathbf{J}|}{J}$ is a unit vector along $\mathbf{J}$, and $\triangle \mathbf{J} = \mathbf{r} \times \mathbf{p}$.

To find coefficients $a_k$ ($k = 0, 1, 2$) and $b_{ik}$ ($i, k = 0, 1, 2$) one needs to substitute $\triangle x_j$, $\triangle x_0$, $\triangle \varphi_j$ into Eqs. (A1), (A2), and (A3) and perform the necessary averaging. Apart from averaging over the part of surface exposed to the flux, one has to average over the angles of precession of $\mathbf{J}$ around $\mathbf{e}_2$ and of $\mathbf{J}$ around $\mathbf{m}$. As a result, one gets for an axially symmetric ellipsoidal grain [Dolginov & Mytrophanov 1976]:

$$a_0 = \frac{c}{J}(2 + s(1 + x_1^2) + g(1 + x_2^2) + 2gs(x_1^2 - x_2^2))$$

$$a_1 = \frac{c}{J^2}(-2 + s(1 - 3x_1^2) - g(1 + x_2^2)$$
$$+ 2gs(1 - 2x_1^2 - 2x_2^2 + 3x_1^2 x_2^2))$$

$$a_2 = \frac{c}{J^2}x_2^2(-2 - s(1 - x_1^2) + g(1 + 3x_2^2)$$
$$+ 2gs(1 - 2x_1^2 - 2x_2^2 + 3x_1^2 x_2^2))$$

$$
\begin{aligned}
b_{00} &= c(1 + s(1 - x_1^2) + g(1 - x_2^2) + gs(1 - 2x_1^2)(1 - x_2^2)) \\
b_{11} &= \frac{c}{J^2}(1 + sx_1^2 + \frac{g}{2}(1 + x_2^2)gsx_1^2(1 - 2x_2^2)) \\
b_{22} &= \frac{c}{J^2}(1 + gx_2^2 + \frac{s}{2}(1 + x_1^2)gsx_2^2(1 - 2x_1^2)) \\
b_{01} &= \frac{c}{J^2}x_1(1 - x_1^2)s(1 + g(1 - x_2^2)) \\
b_{02} &= \frac{c}{J^2}x_2(1 - x_2^2)g(1 + s(1 - x_1^2)) \\
b_{12} &= \frac{c}{J^2}x_1x_2sg(1 - x_1)(1 - x_2)),
\end{aligned}
$$

where

$$
\begin{aligned}
c &= \frac{2}{3}(\langle p^2 \rangle - \langle p_z^2 \rangle)b^2 \\
s &= -\frac{1}{2}(\langle p^2 \rangle - 3\langle p_z^2 \rangle)(\langle p^2 \rangle - \langle p_z^2 \rangle)^{-1} \\
g &= \frac{a^2 - b^2}{2b^2}
\end{aligned}
$$

and $p$ and $p_z$ are the values of the momentum and its $Z_1$ projection transferred in a collision.

## B. Computation of integrals

This double integral and the one in Eq. (14) can be integrated over $\theta_2$ to give

$$
\langle \cos^2 \theta_1 \rangle = C(s, g) \int_0^1 \frac{x^2 \mathrm{d}x}{(1 + sx^2)\sqrt{1 + sx^2 + g}} \tag{B4}
$$

and

$$
C(s, g) \int_0^1 \frac{\mathrm{d}x}{(1 + sx^2)\sqrt{1 + sx^2 + g}} \equiv 1, \tag{B5}
$$

where an evident substitution $x = \cos \theta_1$ is used. Writing

$$
A(1 + sx^2) + B = x^2 \tag{B6}
$$

it is possible to obtain $A = 1/s$ and $B = -1/s$. Therefore

$$
\langle \cos^2 \theta_1 \rangle = C(s, g)\frac{1}{s}\int_0^1 \frac{\mathrm{d}x}{\sqrt{1 + sx^2 + g}} - \frac{1}{s}, \tag{B7}
$$

where the above identity is taken into account. The integral can be calculated [see Gradshtein & Ryzhik 1965, 2.271(4)] to give

$$
\begin{aligned}
i_1 &= \int_0^1 \frac{\mathrm{d}x}{\sqrt{(1+g)+sx^2}} \\
&= \begin{cases} \frac{1}{\sqrt{s}} \ln \frac{\sqrt{s}+\sqrt{1+g+s}}{\sqrt{1+g}} & (s > 0) \\ \frac{1}{\sqrt{-s}} \arcsin \sqrt{-\frac{s}{1+g}} & (s < 0), \end{cases}
\end{aligned}
\tag{B8}
$$

where for uniformity one can also use inverse hyperbolic function for $s < 0$, namely

$$
i_1 = \frac{1}{\sqrt{s}} \operatorname{arcsinh} \sqrt{\frac{s}{1+g}}.
\tag{B9}
$$

To obtain $C(s,g)$, one has to calculate

$$
i_2 = \int_0^1 \frac{\mathrm{d}x}{(1+sx^2)\sqrt{1+sx^2+g}},
\tag{B10}
$$

which can be done by substituting $u = x^2 + s$. The corresponding integral can be easily evaluated [Gradshtein & Ryzhik 1965, 2.224(5)] to give

$$
i_2 = \begin{cases} \frac{1}{2\sqrt{-sg}} \ln \frac{\sqrt{1+s+g}+\sqrt{-sg}}{\sqrt{1+s+g}-\sqrt{-sg}} & (sg < 0) \\ \frac{1}{\sqrt{sg}} \arctan \sqrt{\frac{sg}{1+s+g}} & (sg > 0). \end{cases}
\tag{B11}
$$

The function $C(s,g)$ is equal to $i_2^{-1}$. Using the inverse hyperbolic function the expression for $sg < 0$ is equal

$$
i_2 = \frac{1}{\sqrt{-sg}} \operatorname{arctanh} \sqrt{\frac{-sg}{1+s+g}},
\tag{B12}
$$

which enables one to find $C(s,g)$.

As a result, one obtains for $s < 0$, and $g < 0$

$$
\langle \cos^2 \theta_1 \rangle = \frac{\sqrt{-g} \arcsin \sqrt{-\frac{s}{1+g}}}{s \arctan \sqrt{\frac{sg}{1+s+g}}} - \frac{1}{s},
\tag{B13}
$$

for $s < 0$, and $g > 0$

$$
\langle \cos^2 \theta_1 \rangle = \frac{\sqrt{g} \arcsin \sqrt{-\frac{s}{1+g}}}{s \operatorname{arctanh} \sqrt{-\frac{sg}{1+s+g}}} - \frac{1}{s},
\tag{B14}
$$

for $s > 0$, and $g < 0$

$$\langle \cos^2 \theta_1 \rangle = \frac{\sqrt{-g} \operatorname{arcsinh} \sqrt{\frac{s}{1+g}}}{s \, \operatorname{arctanh} \sqrt{-\frac{sg}{1+s+g}}} - \frac{1}{s}, \tag{B15}$$

for $s > 0$, and $g > 0$

$$\langle \cos^2 \theta_1 \rangle = \frac{\sqrt{g} \operatorname{arcsinh} \sqrt{\frac{s}{1+g}}}{s \, \arctan \sqrt{\frac{sg}{1+s+g}}} - \frac{1}{s}, \tag{B16}$$

which cover all the cases.

## C. Alignment of rotationally hot grains

We have seen in the main body of the paper that if diffusion $\mathbf{J}$ dominates its "leaps", the density of $\mathbf{J}$ in the space of angular coordinates for an ensemble of grains is proportional to $S_n^{1/2}$. Therefore

$$f(\varphi, \phi) = C^{-1} \int_0^{2\pi} \frac{d\psi}{\cos^{1/2} \alpha}, \tag{C17}$$

where

$$C = \int_0^{2\pi} d\psi \int_0^\pi \frac{\sin \varphi \, d\varphi}{\cos^{1/2} \alpha}, \tag{C18}$$

provides an opportunity to obtain the Rayleigh reduction factor for any $\phi$ through numerical integration in Eq(C18)

However in the most important case corresponding to the alignment under Alfvénic perturbations, the calculations can be done analytically. Indeed, $\cos \alpha = \sin \varphi \cos \psi$ and according to Gradshtein & Ryzhik (1965, 3.621(1)),

$$\int_0^{\pi/2} \sin^{3/2-1} x \, dx = 2^{3/2-2} B(3/4, 3/4), \tag{C19}$$

where $B(x, y)$ is beta function. Similarly,

$$\int_0^{\pi/2} \sin^{3/2-1} x \cos^2 x \, dx = 2^{3/2-2} B(3/4, 3/4) - \frac{\pi}{3\sqrt{2} B(9/4, 1/4)}, \tag{C20}$$

where at first $\cos^2 x$ was expressed through $\cos 2x$ and then Gradshtein & Ryzhik (1965, 3.631(8)) was consulted again. Expressing Beta functions through Gamma functions: $B(3/4, 3/4) = \Gamma^2(3/4)/\Gamma(3/2)$ and $B(9/4, 1/4) = \Gamma(1/4)\Gamma(9/4)/\Gamma(5/2)$ and substituting the corresponding values of $\Gamma$ it is possible to obtain

$$\sigma \approx 0.04, \tag{C21}$$

which is much less than the alignment when "leaps" dominate.

Table 1: The analytical expressions for $\langle \cos^2 \theta_1 \rangle$ corresponding to different values of grain non-sphericity $g$ and external anisotropy $s$.

|         | $g < 0$ | $g > 0$ |
|---------|---------|---------|
| $s < 0$ | $\dfrac{\sqrt{-g}\arcsin\sqrt{-\frac{s}{1+g}}}{s\ \arctan\sqrt{\frac{sg}{1+s+g}}} - \dfrac{1}{s}$ | $\dfrac{\sqrt{g}\arcsin\sqrt{-\frac{s}{1+g}}}{s\ \operatorname{arctanh}\sqrt{-\frac{sg}{1+s+g}}} - \dfrac{1}{s}$ |
| $s > 0$ | $\dfrac{\sqrt{-g}\operatorname{arcsinh}\sqrt{\frac{s}{1+g}}}{s\ \operatorname{arctanh}\sqrt{-\frac{sg}{1+s+g}}} - \dfrac{1}{s}$ | $\dfrac{\sqrt{g}\operatorname{arcsinh}\sqrt{\frac{s}{1+g}}}{s\ \arctan\sqrt{\frac{sg}{1+s+g}}} - \dfrac{1}{s}$ |

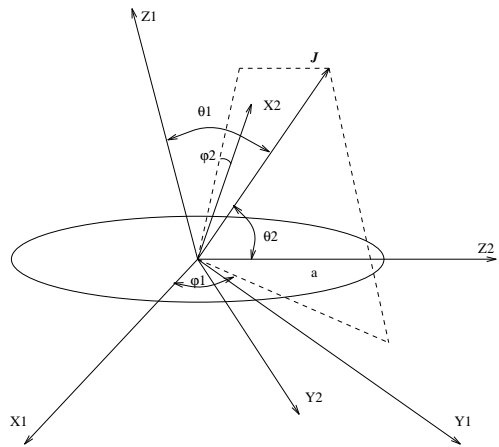

Fig. 1.— $Z_1$ axis of the external or gas reference frame $X_1Y_1Z_1$ is directed along the magnetic field. The internal or grain frame $X_2Y_2Z_2$ is defined so as $Z_2$ coincides with the symmetry axis of the spheroid. $\theta_1$, $\varphi_1$, $\theta_2$, and $\varphi_2$ are the polar angles in the above reference frames.

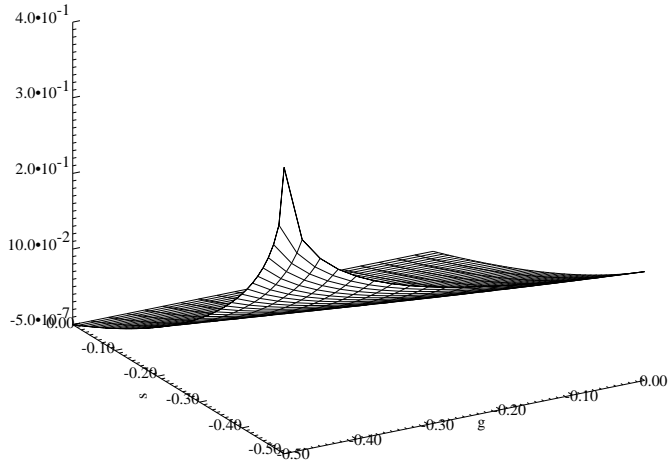

Fig. 2.— The alignment measure $\sigma_J$ for oblate grains $(g < 0)$ for $s < 0$.

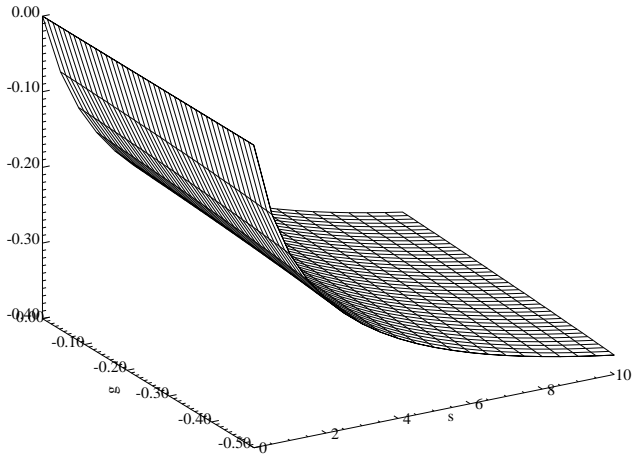

Fig. 3.— The alignment measure $\sigma_J$ for oblate grains $(g < 0)$ for $s > 0$.

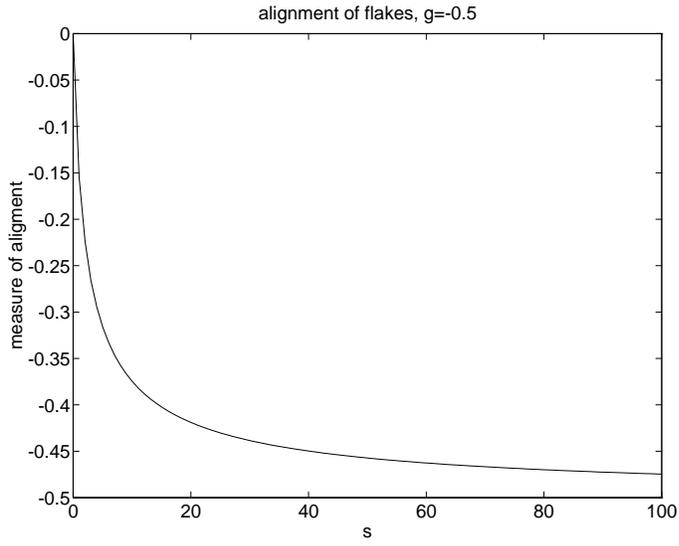

Fig. 4.— The alignment measure $\sigma_J$ for flakes ($g = -0.5$) for $s > 0$.

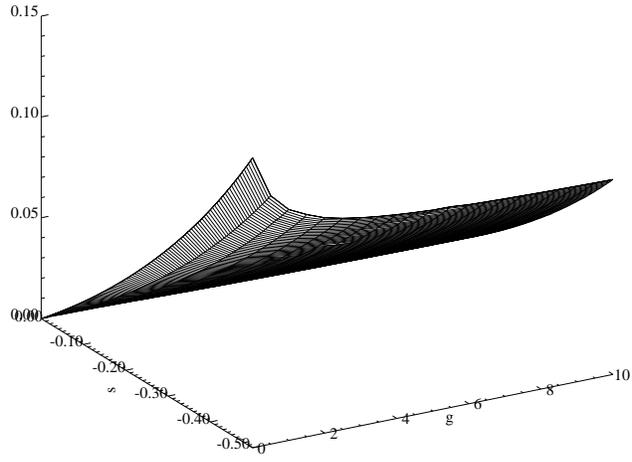

Fig. 5.— The alignment measure $\sigma_J$ for prolate grains $(g > 0)$ for $s < 0$.

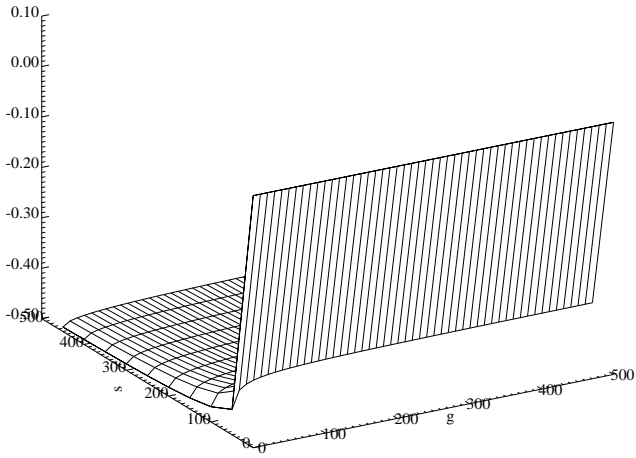

Fig. 6.— The alignment measure $\sigma_J$ for prolate grains ($g > 0$) for $s > 0$.

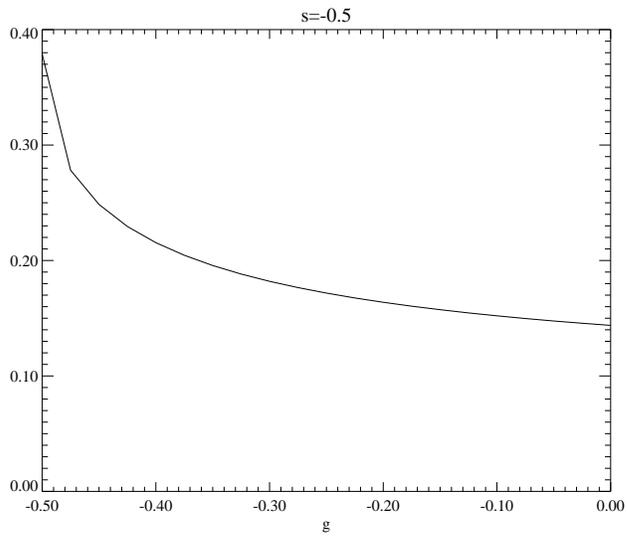

Fig. 7.— The alignment measure $\sigma_J$ for oblate grains ($g < 0$) under Alfvénic perturbations ($s = -0.5$).

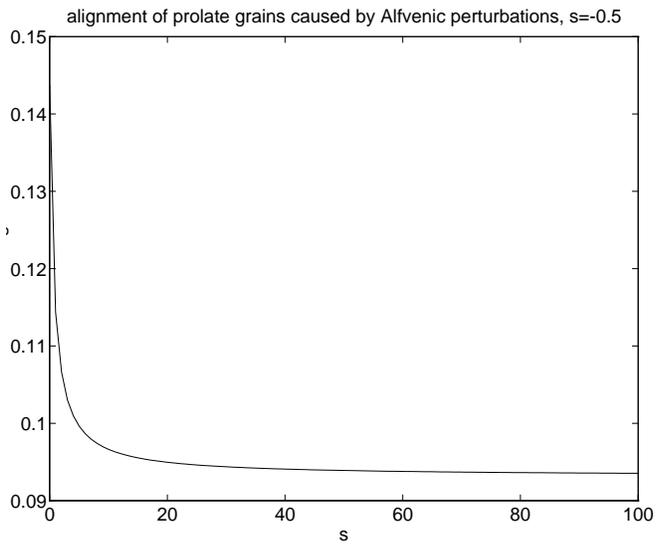

Fig. 8.— The alignment measure $\sigma_J$ for prolate grains $(g > 0)$ under Alfvénic perturbations $(s - 0.5)$.

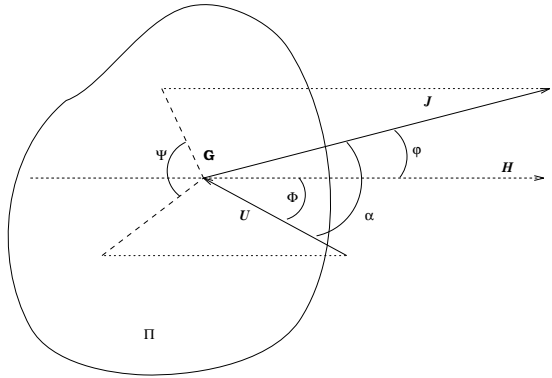

Fig. 9.— Grain at the position **G** is subjected to the flux in the direction given by the relative velocity **u** and its short axis is directed along **J**. The angle $\psi$ is measured in the plane Π which is perpendicular to the direction of magnetic field **H**.